\documentclass{aa}

\usepackage{graphicx}
\usepackage{txfonts}

\begin{document}

\title{Tidally induced bar-like galaxies in simulated clusters}

\author{Ewa L. {\L}okas
}

\institute{Nicolaus Copernicus Astronomical Center, Polish Academy of Sciences,
Bartycka 18, 00-716 Warsaw, Poland\\
\email{lokas@camk.edu.pl}}

\abstract{
One of the scenarios for bar formation in galaxies involves their interaction with a more massive companion. The
stellar component is then transformed from a disk into a bar-like prolate spheroid. I investigated a subsample of 77
bar-like galaxies tidally induced in the cluster environment, selected among the previously studied sample of bar-like
galaxies from the IllustrisTNG simulations. I present six clear, convincing examples of bar-like galaxies formed after
an interaction with a progenitor of a massive brightest cluster galaxy (BCG) and describe the properties of their bars.
For the whole sample, the time of bar formation is strongly correlated with and typically slightly greater than the
time of the pericenter passage. All galaxies are strongly stripped of dark matter and gas, and their rotation is
similarly diminished. A larger pericenter distance typically requires a higher host mass in order to transform the
galaxy, but the interactions show no preference for prograde configurations. The final strength of the bars does not
correlate with the amount of tidal stripping experienced because of the variety of initial properties of the
progenitors and the subsequent evolution over the next pericenter passages. In spite of difficulties in the
interpretation of some cases involving mergers and multiple interactions, the results confirm in the cosmological
context the reality of tidal bar formation in cluster environments previously studied using controlled simulations.}

\keywords{galaxies: clusters: general -- galaxies: evolution -- galaxies: interactions --
galaxies: kinematics and dynamics -- galaxies: spiral -- galaxies: structure  }

\maketitle

\section{Introduction}

\nolinenumbers

Barred galaxies are relatively common in the Universe and easy to detect, for example by visual inspection
\citep{Buta2015}. The main mechanism by which barred galaxies are believed to form is the inherent instability of their
disks, which has been studied extensively using $N$-body simulations, starting with the pioneering works of
\citet{Hohl1971} and \citet{Ostriker1973}. An important factor in the process of bar formation is the presence of a
dark matter halo initially postulated to inhibit the instability, but later shown to play an important role in the
angular momentum transfer \citep{Athanassoula2003}.

Another way that bars form, especially prominent ones, is via tidal interactions with galaxies or clusters. The process
has been studied to date mostly via controlled or idealized simulations of disky galaxies flying by another galaxy
\citep{Noguchi1987, Gerin1990, Miwa1998, Berentzen2004, Lokas2014, Lokas2018} or orbiting in a cluster environment
\citep{Mastropietro2005, Lokas2016}. The tidal force acting on a galaxy in such a configuration is proportional to the
host mass and inversely proportional to the third power of the pericenter distance, and a stronger tidal force leads to
a greater distortion of the galaxy. In agreement with these predictions, numerical experiments demonstrate that for a
massive enough host, a small enough pericenter distance, and preferably a prograde orientation of the disk with respect
to the orbit, a pronounced bar can form right after the pericenter passage. In contrast to bars forming in isolated
disks, which tend to gradually grow in size, the whole structure of the disk is then modified on a short timescale and
the bar is born already mature. This means that not much of the disk component remains in the galaxy and even this is
additionally stripped by tidal forces. The resulting morphology can be described more adequately as bar-like rather
than barred.

The approach based on controlled simulations was adopted in \citet{Lokas2016}, where we studied the formation of a
tidally induced bar in a late-type galaxy similar to the Milky Way, which was placed on different orbits in a
Virgo-like cluster and evolved for 10 Gyr. As a reference case, we also evolved the same model in isolation; it
eventually formed a bar, but very slowly. Tidally induced bars formed on all orbits soon after the first pericenter
passage and survived until the end of the evolution. They appeared earlier, and were stronger and longer, for tighter
orbits. The prediction of this study was that barred galaxies should occur more often and be stronger closer to the
cluster center.

This scenario was later verified in the cosmological context \citep{Lokas2020b} by using the IllustrisTNG simulations
of galaxy formation \citep{Springel2018, Marinacci2018, Naiman2018, Nelson2018, Pillepich2018}, which follow the
evolution of dark matter and baryons solving for gravity and hydrodynamics, and applying additional prescriptions for
star formation, feedback, and other subgrid processes. In that work \citep{Lokas2020b}, I studied the evolution of
well-resolved galaxies in the most massive cluster formed in TNG100 subset of the simulations and demonstrated that the
tidal interaction with the cluster can lead to the formation of tidally induced bars. A few convincing examples of
galaxies were presented in which the formation of a bar coincided with a pericenter passage on an orbit around the
cluster.

In \citet{Lokas2021} I studied the whole population of bar-like galaxies in the IllustrisTNG100 simulation and
investigated their formation scenarios. The sample was selected from the final simulation output (corresponding
to redshift $z=0$) using 6507 subhalos with total stellar masses higher than $10^{10}$ M$_\odot$, which corresponds to
about $10^4$ stellar particles per object, and thus makes the morphological analysis possible. Next, I required
the intermediate to longest axis ratio $b/a$ of the stellar component (within two stellar half-mass radii, $2 r_{1/2}$)
to be lower than 0.6, thus selecting strongly prolate objects. After rejecting some objects whose history could not
be followed to sufficiently early stages to determine the time of bar formation or in which the bar-like shape formed
only very recently (so it could not be considered a stable feature), I obtained the final sample containing 277
bar-like galaxies. I emphasize that these are not bars embedded in disks that have been studied previously using
Illustris and IllustrisTNG simulations \citep{Peschken2019, Rosas2020, Rosas2022, Zhou2020, Zhao2020}. In the latter
works bars were identified in galaxies dominated by rotation-supported disks selected using kinematic decomposition.
The bar-like galaxies studied here were chosen instead only with the axis-ratio condition. Although they may possess a
disk component, their stellar morphology is dominated by a prolate spheroid.

By studying in detail the mass and shape evolution of these galaxies as well as their interactions with other objects I
divided them into three classes based on the origin of the bar and the subsequent history. In galaxies of class A
(comprising 77 objects or 28\% of the sample), the  bar was induced by an interaction with a larger perturber, of mass
in the range $10^{11-14.3}$ M$_\odot$, most often a cluster or group central galaxy, and the galaxies were heavily
stripped of dark matter and gas. In classes B and C (27\% and 45\% of the sample, respectively) the bars were induced
by a merger or a passing satellite, or they were formed by disk instability. Class B galaxies were then partially
stripped of mass, while those of class C evolved without strong interactions, thus retaining their dark matter and gas
content. In spite of the different evolutionary histories, the bars turned out to be remarkably similar in strength,
length, and formation times.

For this work I took a closer look at the 77 bar-like galaxies of class A that had their strong bar induced by an
interaction with a more massive object in the cluster, thus generalizing the previous study \citep{Lokas2020b}, where
only one, the most massive cluster of TNG100, was considered. In Section~2 I present the six most convincing examples
of the formation of tidally induced bar-like objects as a result of an interaction with a cluster-size host. In
Section~3 I describe three more complicated cases with multiple interactions. Section~4 presents another selection of
six cases, in which the strength of the bar grows particularly rapidly, and I try to determine what lies behind this
phenomenon. Section~5 is devoted to the statistical properties of the whole sample, which help us get a clear picture
of the scenario. The discussion follows in Section~6.

\section{Clearest examples of tidally induced bar-like galaxies}

\begin{figure*}[ht!]
\centering
\includegraphics[width=5.6cm]{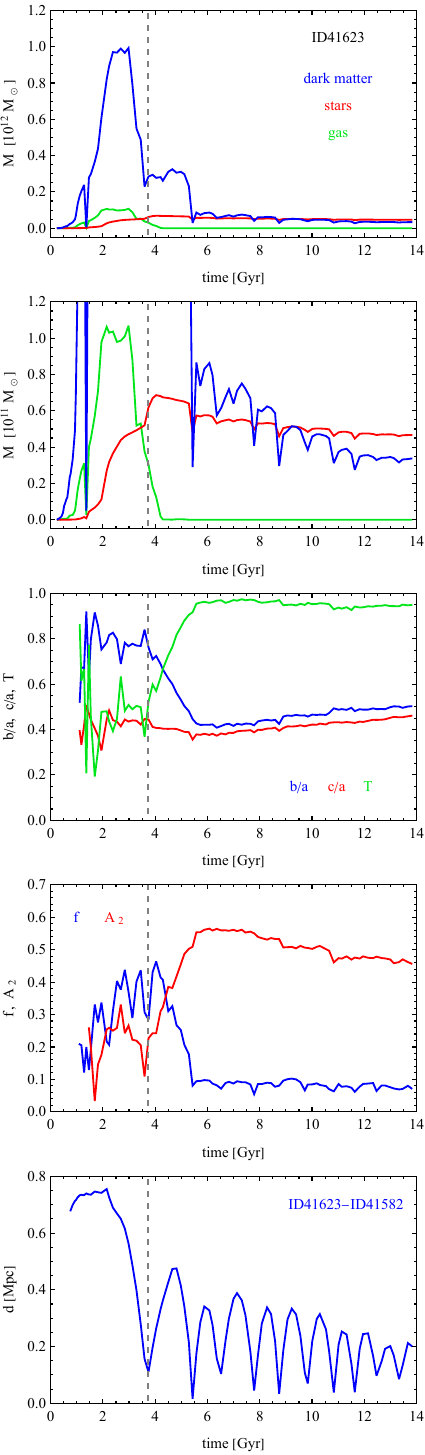}
\includegraphics[width=5.6cm]{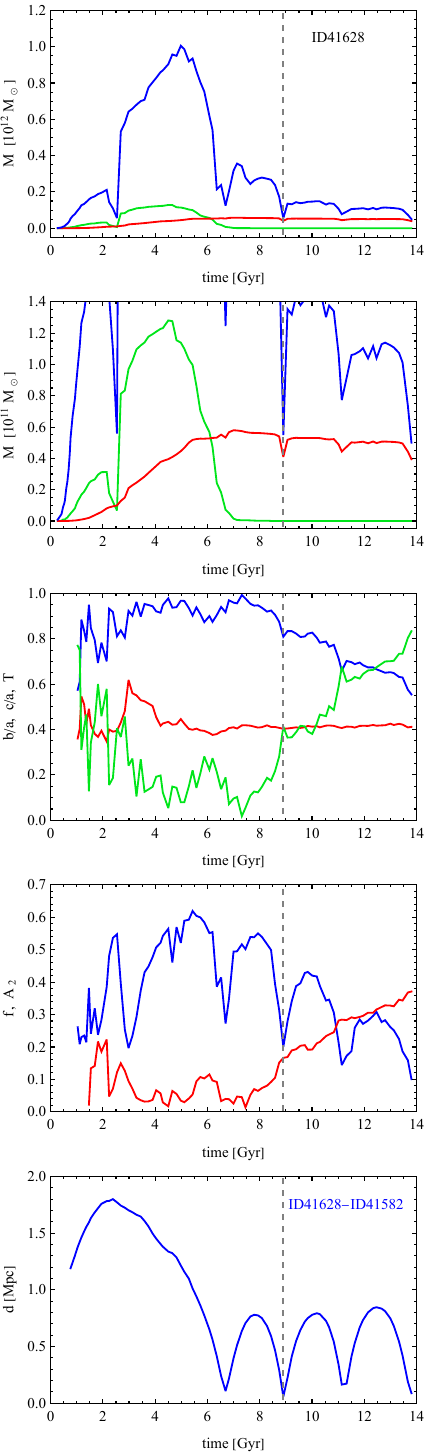}
\includegraphics[width=5.6cm]{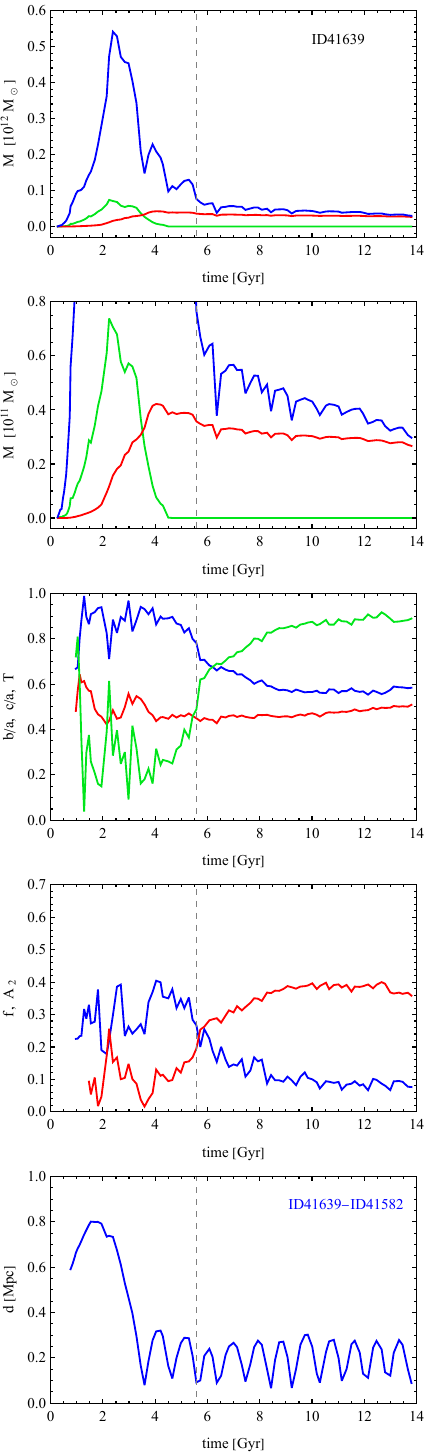}
\caption{Formation of bar-like galaxies in clusters. The columns present the results for different
galaxies, ID41623, ID41628, and ID41639. Upper row: Evolution of the total dark,
stellar, and gas mass shown with the blue, red, and green lines, respectively. Second row: Same masses, but the
vertical scale is adjusted to the stellar and gas mass. Third row: Evolution of three structural properties of the
galaxies, the axis ratios $b/a$ (blue line) and $c/a$ (red), and the triaxiality parameter $T$ (green). Fourth row:
Rotation measure in terms of the fractional mass of stars on circular orbits $f$ (blue) and the bar strength $A_2$
(red). Fifth row: Distance of the galaxy from the cluster BCG. The vertical dashed lines indicate the pericenter at
which the bar was induced.}
\label{evolution1}
\end{figure*}

\begin{figure*}
\centering
\includegraphics[width=5.6cm]{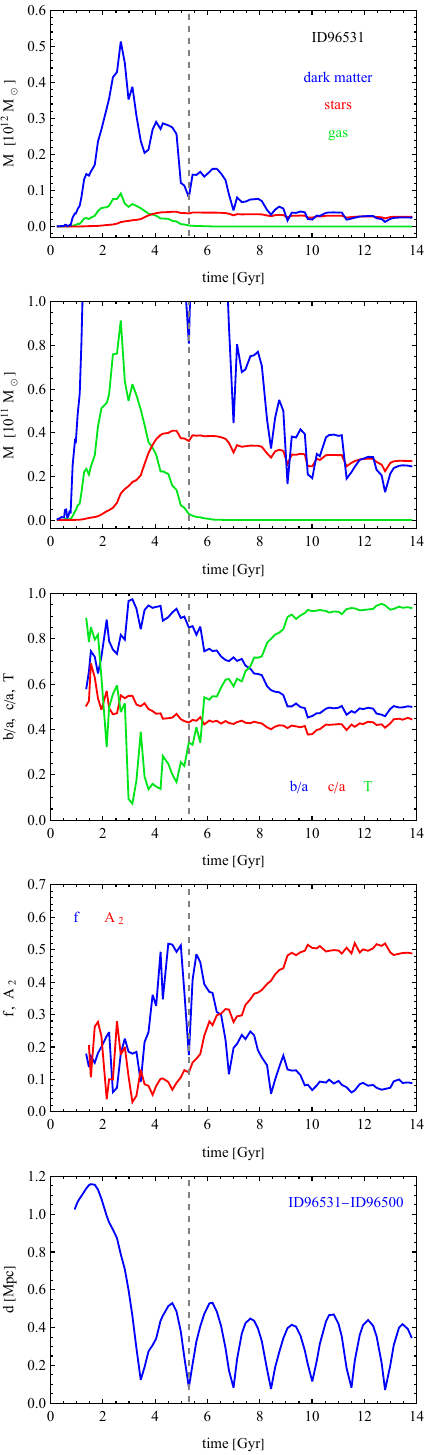}
\includegraphics[width=5.6cm]{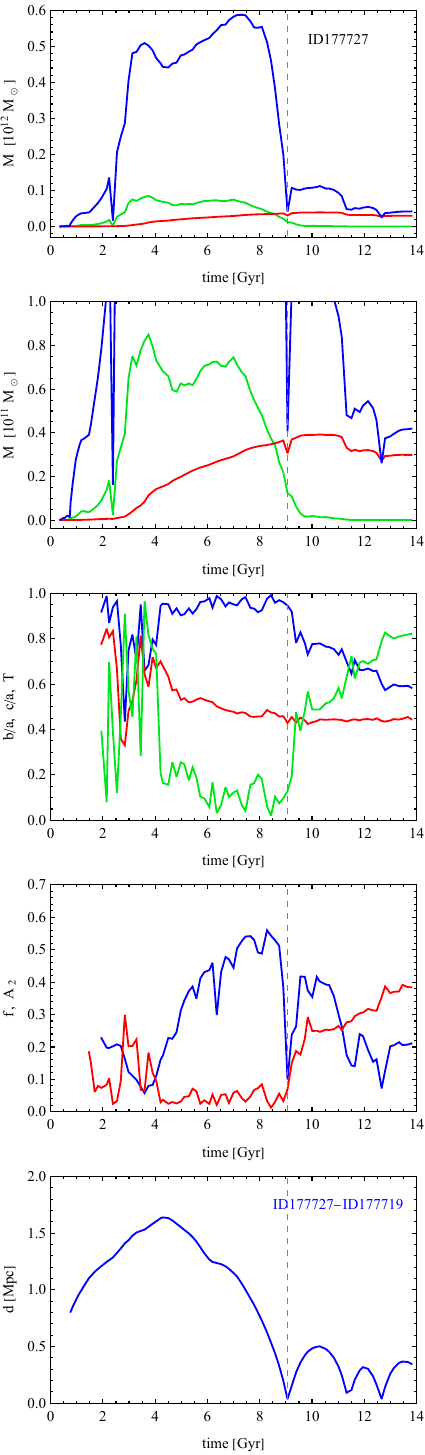}
\includegraphics[width=5.6cm]{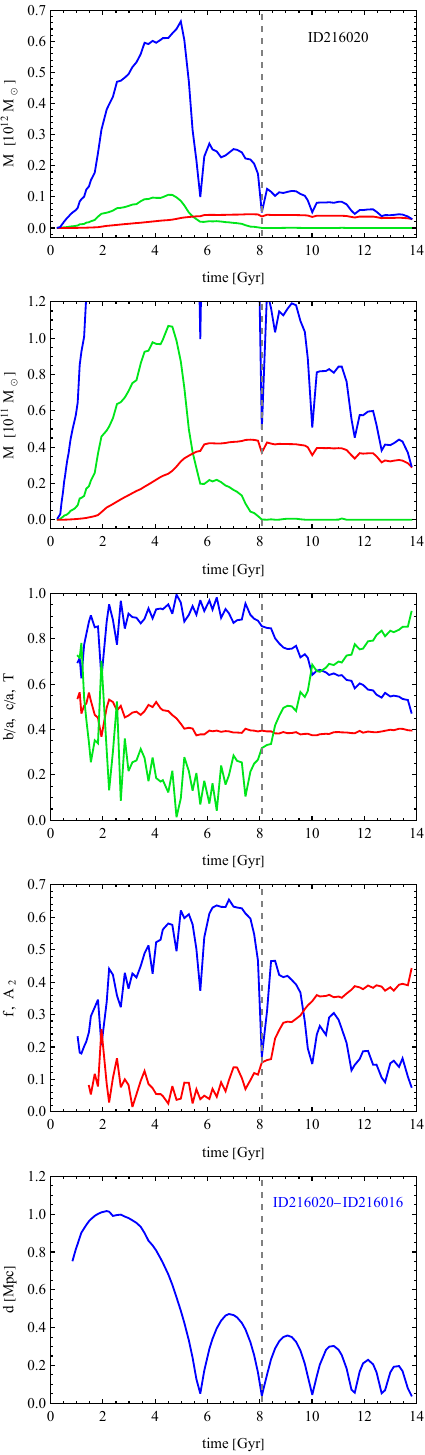}
\caption{Same as Fig.~\ref{evolution1}, but for other examples of bar-like galaxies.}
\label{evolution2}
\end{figure*}

The general properties of the subsample of 77 tidally induced bar-like galaxies differ from the remaining bar-like
galaxies in a few respects. As discussed in \citet{Lokas2021}, due to strong tidal stripping they are much less (or not
at all) dominated by dark matter. They all have fractions of dark to stellar total mass $M_{\rm dm}/M_* < 10$ and in
some of them the dark matter mass is even a few times lower than the stellar mass, while similar galaxies in
low-density environments typically have $M_{\rm dm}/M_* \sim 50$. By the end of the evolution, they are also completely
devoid of gas (except for one object with a very small gas fraction), which was lost due to ram-pressure stripping by
the hot intracluster gas, a process known to play a crucial role in the evolution of galaxies in clusters
\citep{Gunn1972, Boselli2022}. As a result of tidal truncation, their stellar components are more compact than in the
other subsamples, with smaller stellar half-mass radii, they are completely quenched and redder than the rest. The
study also showed that they formed from disks with smaller amount of rotation.

Figures~\ref{evolution1} and \ref{evolution2} present the six most convincing examples of the formation of bar-like
galaxies resulting from interactions with more massive objects. These are particularly clear cases, where the
morphological transformation immediately follows a strong interaction with an object easily identifiable as the central
galaxy of a cluster. The properties of the galaxies (subhalos) were extracted from the publicly available simulation
data for the TNG100 suite of the IllustrisTNG project, as described by Nelson et al. (2019). The galaxies are
identified by their corresponding subhalo identification numbers at the final simulation output ($z=0$).

The upper panels of the figures show the evolution of the total mass of the galaxies in three components: dark matter
(blue), stars (red), and gas (green). One can immediately see that after the initial growth, the dark matter mass drops
suddenly and very strongly, which is a characteristic signature of tidal stripping. The second row of panels shows the
same data with the scale adjusted so that the stripping of stars and gas is more discernible. While the stars are
stripped only weakly and gradually, the gas is lost very quickly and completely.

The third row of panels in Figs.~\ref{evolution1} and \ref{evolution2} illustrates the evolution of the key property of
this analysis, namely the shape of the stellar component of the galaxies in terms of the intermediate-to-major ($b/a$)
and short-to-major ($c/a$) axis ratios, and the triaxiality parameter $T = [1-(b/a)^2]/[1-(c/a)^2]$ (the blue, red, and
green lines, respectively). The axis ratios were estimated from the eigenvalues of the mass tensor of the stellar
component included within two stellar half-mass radii, $2 r_{1/2}$. For all galaxies, the thickness of the galaxy, as
measured by $c/a$, remains approximately constant during the later stages of evolution, while $b/a$ decreases and $T$
increases signifying a transition from the oblate shape of the disk to the prolate shape of the forming bar.

The fourth row of panels shows the evolution of two more parameters related to the shape and kinematics. The rotation
parameter $f$ (blue line) is defined as the fractional mass of all stars with circularity parameter $\epsilon > 0.7$,
where $\epsilon=J_z/J(E),$ and $J_z$ is the specific angular momentum of the star along the angular momentum of the
galaxy, while $J(E)$ is the maximum angular momentum of the stellar particles at positions between 50 before and 50
after the particle in question in a list where the stellar particles are sorted by their binding energy
\citep{Genel2015}. One can see that during the later stages of the evolution $f$ decreases from values around 0.4 or
higher characteristic of disks down to values below 0.2 characteristic of objects no longer supported by rotation.

The red lines in the fourth-row panels present the evolution of the commonly used measure of the strength of the bar
\citep{Athanassoula1986, Athanassoula2002, Athanassoula2013} in the form of the $m=2$ mode of the Fourier decomposition
of the surface density distribution of stellar particles projected along the short axis. It is given by $A_2 (R) = |
\Sigma_j m_j \exp(2 i \theta_j) |/\Sigma_j m_j$, where $\theta_j$ is the azimuthal angle of the $j$th star, $m_j$ is
its mass, and the sum goes up to the number of particles in a given radial bin. The single-value measurements presented
here were made using all stars within two stellar half-mass radii, $2 r_{1/2}$. As expected, the values of $A_2$
increase at about the same time as the triaxiality, almost tracing its behavior, and rapidly cross the threshold value
of 0.2, which can be adopted as a signature of strong bar formation.

\begin{figure*}
\centering
\includegraphics[width=5.5cm]{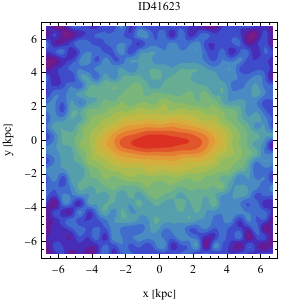}
\includegraphics[width=5.5cm]{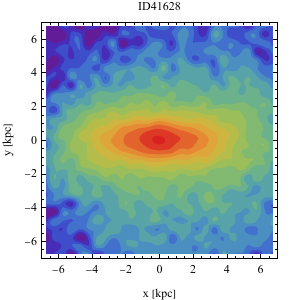}
\includegraphics[width=5.5cm]{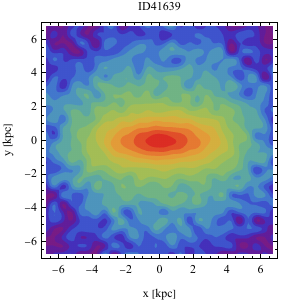}\\
\vspace{0.1cm}
\includegraphics[width=5.5cm]{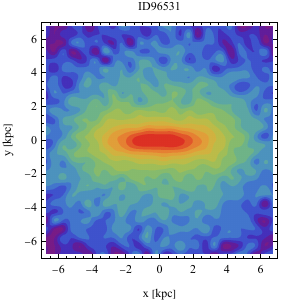}
\includegraphics[width=5.5cm]{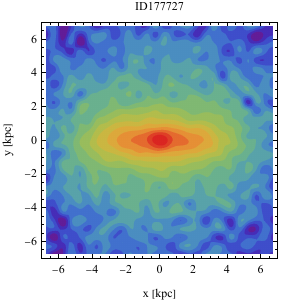}
\includegraphics[width=5.5cm]{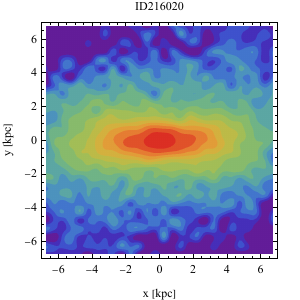}
\caption{Surface density distribution of the stellar components of the six selected bar-like galaxies in the face-on
view at the present time. The surface density, $\Sigma,$ is normalized to the central maximum value in each case, and
the contours are equally spaced in $\log \Sigma$.}
\label{surden}
\end{figure*}

Looking for neighboring objects at the time of bar formation one can identify those that could induce the bar via tidal
interactions. Estimating the tidal force exerted by the neighbors, it is then usually possible to identify the one
which likely had the largest effect on the galaxy forming the bar. In the case of the galaxy ID41623 (left column of
Fig.~\ref{evolution1}) the perturber turns out to be the most massive progenitor of the brightest cluster galaxy (BCG)
ID41582. The galaxy had its first pericenter of $r_{\rm peri} = 110$ kpc around this host at $t = 3.7$ Gyr ($z =
1.7$), and the bar formed ($A_2$ crossed the threshold of 0.2) around the same time. Although this is rather early in
cosmic history, the host was already quite massive at that time, with $M_{\rm host} = 4.2 \times 10^{13}$ M$_\odot$.
The barred galaxy then remained on a tight orbit around this BCG, surviving until the present in spite of many
pericenter passages (lower left panel of Fig.~\ref{evolution1}) and strong mass loss. The bar mode increased strongly
for some time until about 6 Gyr and decreased gradually later on.

The galaxy ID41628 (middle column of Fig.~\ref{evolution1}) formed its bar much later, at $t = 10.2$ Gyr ($z = 0.3$)
following its second pericenter of $r_{\rm peri} = 58$ kpc at $t = 8.9$ around the same BCG, then almost fully formed
and more massive, with $M_{\rm host} = 1.2 \times 10^{14}$ M$_\odot$. The orbit of the galaxy around the cluster was
less tight than in the previous case, with only four pericenter passages in total, which resulted in less dramatic loss
of dark matter and rotation, while the bar continued to grow in strength until the present time.

The third galaxy of this set, ID41639 (right column of Fig.~\ref{evolution1}), formed its bar at $t = 5.7$ Gyr, right
after the third pericenter with $r_{\rm peri} = 92$ kpc, which occurred at $t = 5.6$ Gyr, on a very tight orbit around
the same BCG, which then had the mass $M_{\rm host} = 6.9 \times 10^{13}$ M$_\odot$. In this case the mass and rotation
were lost very efficiently, but the bar retained its strength for the last few gigayears.

The three galaxies in Fig.~\ref{evolution1} show quite different behavior of $A_2$ in the later stages of the evolution.
While for ID41623 the bar mode decreased, for ID41628 it grew until the present time, and for ID41639 it flattened and
remained roughly constant in spite of many pericenter passages. The origin of this phenomenon is discussed in
Section~5.

Figure~\ref{evolution2} shows three further examples of bar-like galaxies formed as a result of the interaction with a
massive BCG, with $M_{\rm host} > 10^{13}$ M$_\odot$, but this time a different one for each galaxy. The formation of
the bar occurred after the first or the second pericenter passage around the host and was accompanied by strong mass and
rotation loss. All the six examples presented in Figs.~\ref{evolution1} and \ref{evolution2} are particularly clear
cases of the tidal formation of bar-like galaxies because the formation of the bar coincided with or followed the
pericenter passage and the host causing the morphological transformation was easily identifiable. In addition, the
mergers the galaxy experienced during its formation stopped well before it started to orbit the cluster, which means
that they did not modify the structure of the galaxy any more and speaks in favor of the tidal interaction as the
dominant factor shaping their morphology.

Let us now have a closer look at the bars of the six selected galaxies. Figure~\ref{surden} shows their images in terms
of the surface density distribution of the stars in the face-on view, in which the bar is best visible. The images show
the galaxies at the last simulation output corresponding to the present time and demonstrate that the bars, although
formed at very different times, survive until the end of the evolution, and are thus very stable structures. In
addition, the images show that most of the stellar component is in the form of the bar, namely there is essentially no
disk left. The reason for this is the formation mechanism, which tends to transform the whole galaxy from a disk to a
bar at the pericenter passage, but also the tidal stripping, which removes the remaining outer stars as the galaxy
continues to orbit the cluster after bar formation.

\begin{figure}
\centering
\includegraphics[width=7.3cm]{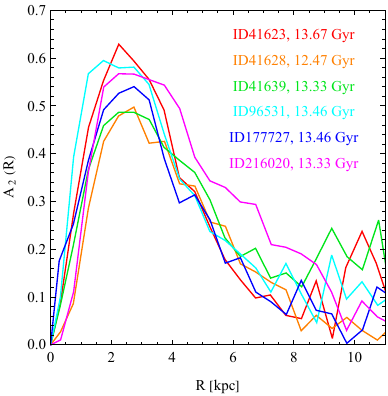}
\caption{Profiles of bar mode, $A_2 (R)$, for the six selected galaxies, at their respective last apocenters on the
orbit around the cluster. Measurements were carried out in bins of $\Delta R = 0.5$ kpc.}
\label{a2profiles}
\end{figure}

Figure~\ref{a2profiles} presents the full profiles of the bar mode $A_2 (R)$ for the six selected galaxies rather than
the single-value measurements shown in Figs.~\ref{evolution1} and \ref{evolution2}. They were measured at the last
apocenter of the orbit of each galaxy around its host rather than at the last simulation output. The reason for this
choice is that at pericenters (where half of the selected galaxies happen to end their evolution) the $A_2 (R)$
profiles are strongly perturbed and increase monotonically up to unity \citep{Lokas2014}. Instead, at apocenters of the
orbits, the galaxies regain their equilibrium and can be considered as relaxed objects. As demonstrated by
Fig.~\ref{a2profiles}, the bar modes then behave in a normal way, characteristic for typical barred galaxies: they
increase with radius, reach a maximum, and then decrease to zero again. The value of the maximum $A_{\rm 2,max}$ can be
used as a measure of the strength of the bar, while the radius at which $A_2 (R)$ drops to half the maximum is useful
as an estimate of the bar length. As can be seen from Fig.~\ref{a2profiles}, the length of the bar is around 5-6 kpc
for all galaxies.

\begin{figure}
\centering
\hspace{0.4cm}
\includegraphics[width=3.5cm]{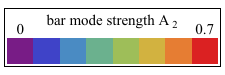}\\
\vspace{-0.5cm}
\includegraphics[width=9.cm]{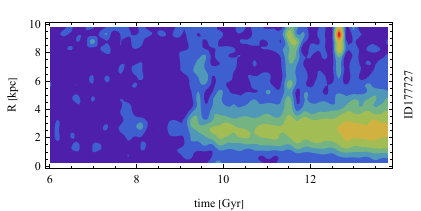}
\caption{Evolution of the profile of the bar mode, $A_2 (R)$, over time for galaxy ID177727.}
\label{a2modestime}
\end{figure}

The evolution of the profiles of the bar mode can be fully appreciated in two-dimensional images showing $A_2(R, t)$
as a function of both radius and time. Such images are difficult to produce, however, for the sample of galaxies
studied here because at pericenters, not only are the $A_2 (R)$ distorted, but the galaxies are strongly truncated and
have no stars bound to them in the outer bins. For this reason, in Fig.~\ref{a2modestime} I show such a plot for just
one galaxy, ID177727, which had only three pericenter passages and was relatively weakly affected by tidal forces. This
one image is sufficient to illustrate the general idea of the $A_2 (R, t)$ evolution. Comparing with the plot of the
orbit for this galaxy in the lower middle panel of Fig.~\ref{evolution2}, one can see that it had three pericenter
passages around its host, at $t = 9.1, 11.3$, and 12.7 Gyr. These are exactly the times where the $A_2 (R, t)$ values
show significant increase in Fig.~\ref{a2modestime}, not only near the center, but out to large radii. These increases
correspond to tidal distortions of the galaxy into an elongated shape. Between pericenters the elongation in the outer
parts disappears, but remains in the inner part, in the form of the bar. Once formed at the first pericenter passage,
the bar is stable and even increases its strength at the third pericenter.

\begin{figure*}
\centering
\includegraphics[width=5.6cm]{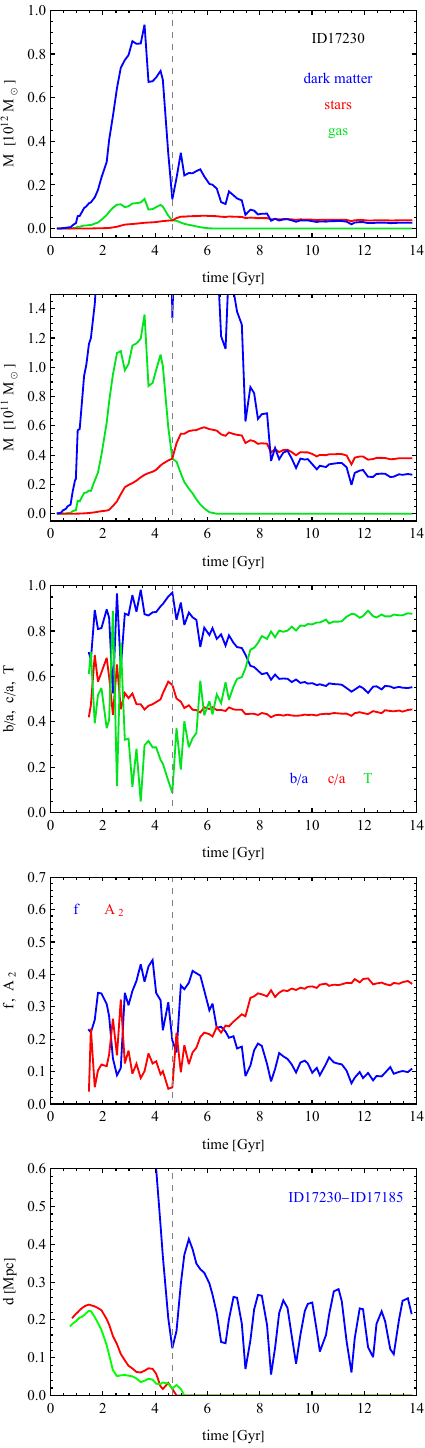}
\includegraphics[width=5.6cm]{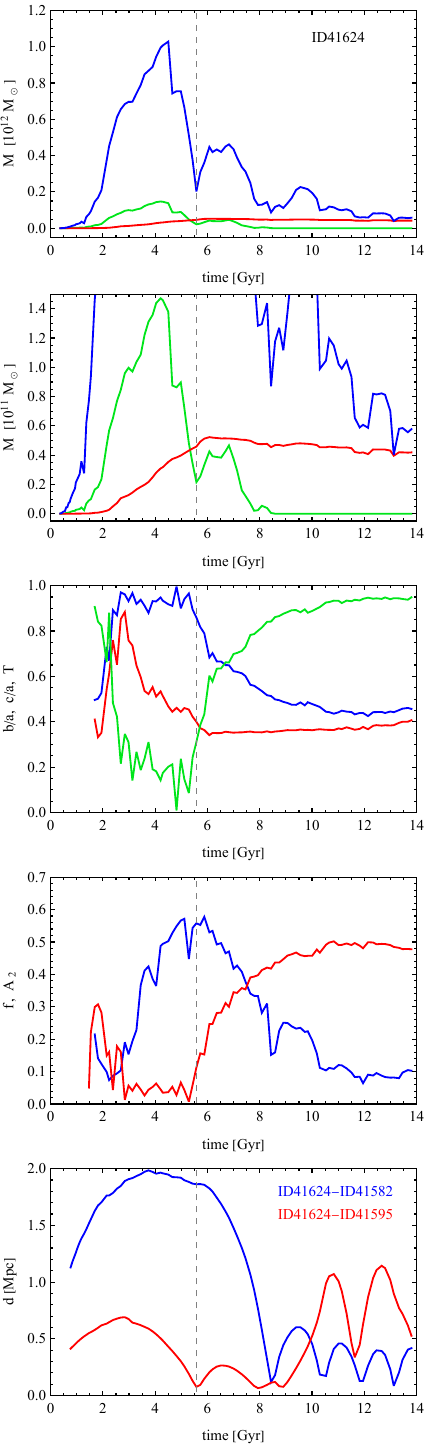}
\includegraphics[width=5.6cm]{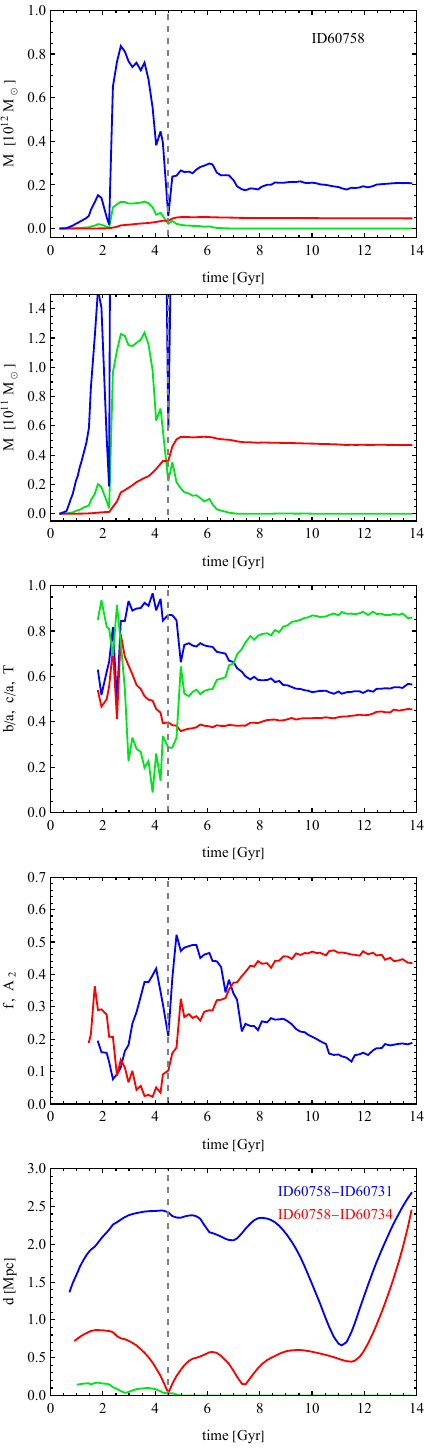}
\caption{
Same as Fig.~\ref{evolution1}, but for other examples of bar-like galaxies with a more complicated history.
In the last row of panels, in addition to the distance of the galaxy from the BCG, plotted as before with the
blue line, the red and green lines show the distance to other perturbers.}
\label{evolution3}
\end{figure*}

\section{More complicated cases with multiple interactions}

Contrary to the clearest cases presented in Section~2, where a disky galaxy interacts with the most massive
progenitor of the BCG and forms a bar, many of the galaxies in the sample studied here undergo a much more complicated
evolution involving multiple interactions with many galaxies at the same time and mergers. There is no discernible
pattern to those interactions in a sense that every galaxy evolves differently. To give an idea of the variety of
possibilities, in this Section I present three examples of such cases. Figure~\ref{evolution3} illustrates the
evolution of three galaxies in a way similar to Figs.~\ref{evolution1} and \ref{evolution2}, except that in the panels
of the lowest row, in addition to the blue lines showing the distance of the galaxy to its BCG host, I also included
lines depicting the distance from other perturbers, with which a given galaxy interacted.

The galaxy ID17230 (left column of Fig.~\ref{evolution3}) formed its bar soon after the first interaction with
the most massive progenitor of BCG ID17185, so in a way similar to the clear cases described in the previous section.
At the time of the interaction the host had the mass of $4.6 \times 10^{13}$ M$_\odot$ and the pericenter distance was
rather small, of 125 kpc, so the tidal force seems to have been strong enough to induce a bar. It should be noted that
at this time the galaxy also interacted with three other galaxies more massive than itself, two of which later merged
with the BCG. The masses of these galaxies were, however, two orders of magnitude lower than the main progenitor of the
BCG, so their effect was probably weak. More importantly, at the time of the major interaction, at $t = 4.7$ Gyr, the
galaxy also experienced two mergers with satellites of significant mass above $10^{10}$ M$_\odot$ (before the
interaction). As a result, the bar mode $A_2$ (fourth panel from top) did not grow steadily after the pericenter, but
experienced some variation before reaching the threshold of 0.2. In this case, the mergers seem to have slowed down bar
formation because the time delay between the interaction with the BCG and the bar formation was as large as 1 Gyr.

In the case of galaxy ID41624 (middle column of Fig.~\ref{evolution3}), the bar was induced as a result of a
strong interaction with a massive galaxy ID41595 at $t = 5.6$ Gyr. The perturber had a mass of $5.2 \times 10^{12}$
M$_\odot$ at the time of the interaction, and the pericenter distance was 76 kpc. The bar-like galaxy remained on an
orbit around this host until about $t = 8.5$ Gyr when it became a satellite of a more massive BCG ID41582. It remained
on a tight orbit around this new host, losing mass and making its bar even a bit stronger. The initial perturber also
interacted with the BCG and lost mass, never becoming a BCG itself, and its distance with respect to the bar-like
galaxy later increased. In this case there were no significant mergers at the time of the interaction or later.

The evolution of the third galaxy ID60758, shown in the right column of Fig.~\ref{evolution3}, combines the
history of the two objects described above. In this case the interaction with BCG ID60731 is weak and happens late in
the evolution, and in fact the bar-like galaxy ends up in the outskirts of the cluster. The bar was induced much
earlier as a result of a strong interaction at $t = 4.5$ Gyr with galaxy ID60734 of high mass, $1.7 \times 10^{13}$
M$_\odot$ at that time (but losing some of it later), with a pericenter distance of only 34 kpc. However, right after
this interaction, at the time of bar formation ($t = 5$ Gyr) the galaxy ID60758 also experienced a significant merger
with a satellite of mass above $10^{10}$ M$_\odot$, which may have helped the bar to form, because the increase of the
bar mode was particularly strong exactly at that time. Because of the weak interactions with the original perturber and
the BCG in the later stages of the evolution, this bar-like galaxy experienced a relatively small mass and rotation loss
compared to the other cases shown in Fig.~\ref{evolution3}.

\section{Cases with strongest growth of the bar mode}

\begin{figure*}
\centering
\includegraphics[width=16cm]{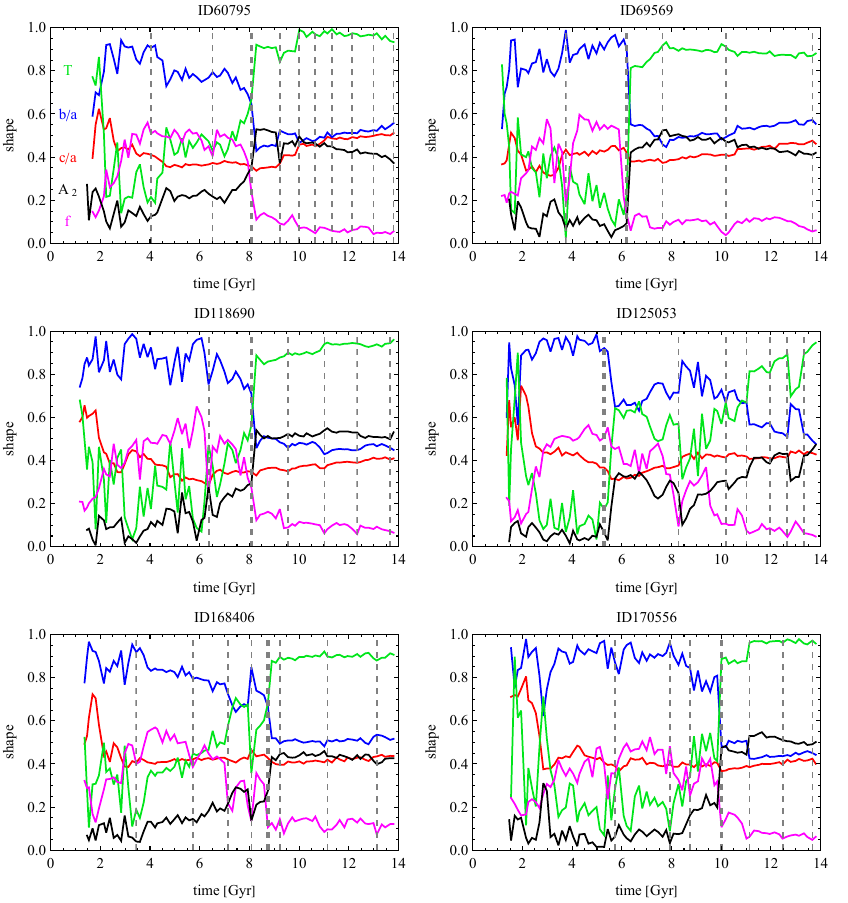}
\caption{Cases of bar-like galaxies showing the strongest increase in the bar strength at one of the pericenters. Each
panel shows results for a galaxy with ID given at the top of the plot. The lines display the evolution of the axis
ratios $b/a$ (blue line), $c/a$ (red), the triaxiality $T$ (green), the bar mode $A (<2 r_{1/2})$ (black), and the
rotation parameter $f$ (magenta). The vertical dashed lines indicate pericenter passages;  the thicker one marks
the interaction causing the strongest increase in the bar mode.}
\label{strongbars}
\end{figure*}

In this Section I propose to examine a few more cases of bar-like galaxy evolution that involve instances
of especially strong growth of the bar mode, not necessarily related to the bar formation per se, that is the crossing
of the threshold $A_2 = 0.2$. In addition to the cases presented in the previous sections, these examples also serve to
illustrate the complications one encounters when trying to interpret different configurations of galaxies in relation
to bar evolution.

Figure~\ref{strongbars} illustrates the evolution of the most important properties of the selected galaxies related to
their morphology. Each panel shows for a given galaxy the axis ratios $b/a$ (blue line), $c/a$ (red), the triaxiality
$T$ (green), the bar mode $A (<2 r_{1/2})$ (black), and the rotation parameter $f$ (magenta) as a function of time. The
pericenter passages a given galaxy underwent are marked as vertical dashed lines, with the pericenter causing the
strongest increase in $A_2$ highlighted with a thicker line.

The galaxy ID60795 (upper left panel of Fig.~\ref{strongbars}) started forming its bar early on, around 4 Gyr when it
experienced its first passage around a massive galaxy ID60734, although not a BCG progenitor. At this time it also had
some last minor mergers. Nominally, it formed a bar at 7.1 Gyr, after the second pericenter that took place at 6.5 Gyr.
But only at the third pericenter, at 8.1 Gyr, did the galaxy experience its strongest growth of the bar mode. As the
orbit tightened, the bar strength was subject to further evolution with the bar mode decreasing and then increasing
again at the fourth and fifth pericenters.

In the case of galaxy ID69569 (upper right panel of Fig.~\ref{strongbars}) the pericenter causing the bar formation
was also the one when the increase of the bar mode was the strongest and the galaxy was not perturbed by any
mergers. The bar was probably induced by an interaction with a progenitor of BCG ID69507 at 6.2 Gyr, but at
the same time the galaxy also experienced a close flyby of another galaxy exerting a tidal force of similar magnitude.

The case of galaxy ID118690 (middle left panel of Fig.~\ref{strongbars}) is similar to ID60795 in a sense that the
strongest increase of the bar mode took place one pericenter later than the one that led to the bar formation. This
strong increase coincided with the second passage around the progenitor of BCG ID118679. The problematic circumstances
of this galaxy include the fact that the first passage was accompanied by interactions with other smaller galaxies and
some minor mergers so that the bar growth was not monotonic.

The case of ID125053 (middle right panel of Fig.~\ref{strongbars}) is perhaps the most interesting of the six examples.
For this galaxy, the strongest growth of the bar mode took place well before the formation of a stable bar, at the
first pericenter passage at 5.3 Gyr around a progenitor of BCG ID125027. A bar formed, but was subsequently weakened by
the last merger at 7 Gyr and only rebuilt into a stable bar after the next pericenter passage at 8.3 Gyr. Although
remaining above the threshold of 0.2, the bar mode showed considerable variations at the next pericenters, including a
strong drop and then increase at the penultimate and the last pericenters. This is an example of the strong
variation of the bar strength related to the orientation of the bar at the pericenter, discussed in the next section.

The galaxy ID168406 (lower left panel of Fig.~\ref{strongbars}) started to form its bar early, at 3.5 Gyr, after a
passage close to one of the progenitors of BCG ID168390 and the last merger. The bar eventually formed after the fourth
pericenter passage at 8.1 Gyr, but strongly increased its strength at the following very tight pericenter at 8.7 Gyr.
The galaxy was then ejected on a wider orbit around this host probably as a result of interactions with other
massive galaxies of the cluster, but the strength of the bar remained unchanged until the present time.

For ID170556 (lower right panel of Fig.~\ref{strongbars}) the bar started to form at the third pericenter around the
progenitor of BCG ID170540 at 8.7 Gyr, when it also had its last merger. It was fully formed only after the
fourth passage around this host, at 10 Gyr, when it also experienced its strongest growth, supported by
close interactions with other galaxies of the cluster. The bar was further enhanced at the fifth pericenter passage at
11.1 Gyr.

\begin{figure*}
\centering
\includegraphics[width=16.3cm]{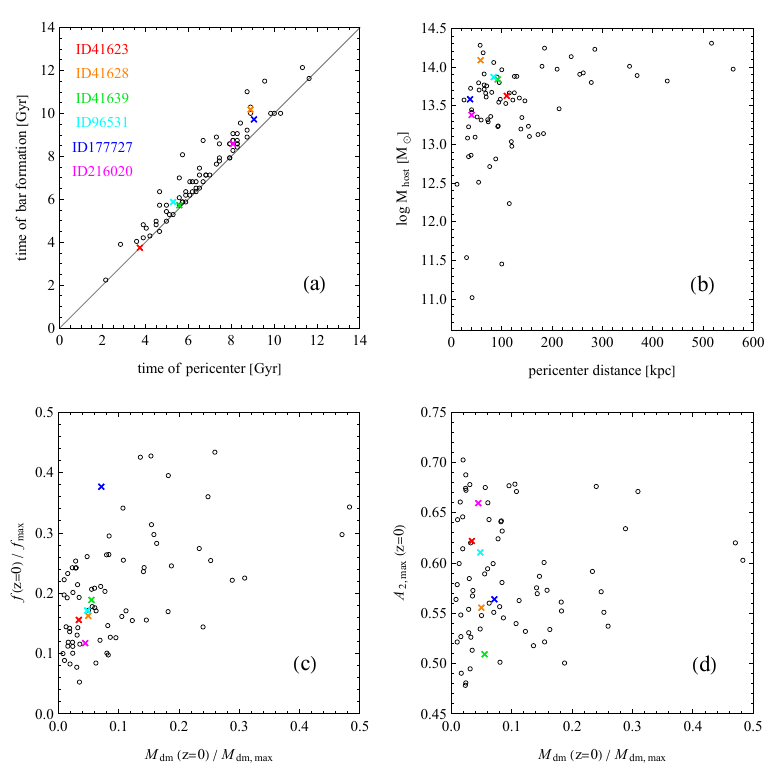}
\caption{
Properties of the whole sample of 77 tidally induced bar-like galaxies.
(a) Time of bar formation as a function of time of pericenter. The diagonal line indicates the
equality between the values.
(b) Host mass as a function of pericenter distance.
(c) Fraction of remaining rotational support as a function of the fraction of remaining dark matter mass.
(d) Strength of the bar $A_{2, \rm{max}}$ as a function of the fraction of remaining dark matter mass at the
present time.
In all panels the colored crosses indicate the values for the selected six galaxies described in detail in Section~2,
while the open black circles show the values for the remaining galaxies.
}
\label{properties}
\end{figure*}

\section{Properties of the sample}

Having described different cases of bar-like galaxies formed by tidal interactions, let us now look at the
properties of the whole sample of 77 tidally induced bar-like galaxies. Two of the interesting properties to compare are
the time of the pericenter passage and the time of bar formation. Let me recall, that the time of bar formation is
defined as the moment when the value of the $A_2$ (measured within $2 r_{1/2}$) crosses the threshold of 0.2 and
remains above it until the end of the evolution. Since the tidal forces are largest at the pericenter passage, the time
of bar formation should coincide or be a little later than the time of the pericenter if there is a causal relation
between the two events. It should be noted that the time step between the subsequent outputs available for study in the
IllustrisTNG simulations is in the range 0.1-0.2 Gyr so this is also the accuracy of the estimates of the two values.

Figure~\ref{properties}a shows the time of bar formation as a function of the time of pericenter for 77 bar-like
galaxies in the sample. The colored crosses indicate the values for the selected six galaxies described in detail in
Section~2 and the values for the rest of the galaxies are shown with the open black circles. The black diagonal line
indicates the equality between the two parameters. The data show that the points are indeed strongly correlated (with
the correlation coefficient of 0.96) and almost all lie on or above the line so that the time of bar formation is the
same or later than the time of the pericenter. In particular, all the colored crosses show this behavior.

There are two galaxies, however, for which the bar forms a little before the pericenter passage. In these cases the bar
started to grow already before the interaction and $A_2$ happened to reach the threshold of 0.2 before the pericenter
passage, after which it further increased, which means that the interaction had an important role in bar
formation. One should also note that in some cases the difference between the pericenter time and the bar formation time
can exceed 2 Gyr. In these cases the bar formation is slow due to a weak interaction or minor mergers. An example
of the delay caused by mergers is provided by the case of galaxy ID17230 described in Section~3 (Fig.~\ref{evolution3},
left column), where the tidal interaction coincided with two mergers delaying the bar formation by 1 Gyr with respect
to the time of the pericenter passage. The galaxy with the longest delay of 2.3 Gyr, ID96524, started forming a bar
after a weak interaction with BCG ID96500, but also interacted with a few other less massive galaxies of the cluster
at that time. The galaxy ID250397, with a similar delay, started to form a bar after a mild interaction with a not very
massive BCG ID250390, but also interacted with other galaxies and experienced mergers. In addition, at the time of
the interaction, the galaxy was still gas-rich, which may have additionally slowed down the morphological
transformation.

Figure~\ref{properties}b illustrates the properties of the sample in terms of the host masses as a function of the
pericenter distances measured at the time of interaction. It should be noted that the pericenter values are the upper
limits of the real ones because of the 0.1-0.2 Gyr difference in time between subsequent outputs of the simulation. The
relation between these parameters is clear: smaller perturber masses require smaller pericenters to exert a
sufficient tidal force on the galaxies. In particular, larger pericenters, above 200 kpc, require host masses of
about $10^{14}$ M$_\odot$ in order to induce the bar. Interestingly, there are three cases, where the perturber
mass was below $10^{12}$ M$_\odot$. For these galaxies, the formation of the bar could be directly associated with a
flyby of a companion of a similar mass, rather than a BCG or its massive progenitor, in spite of the fact that the
galaxies also ended up as cluster members. Two of these cases (ID96526 and ID96527) were described in detail in
\citet{Lokas2023} as undergoing preprocessing inside a small group infalling into a cluster. The six selected example
galaxies discussed in Section~2 (colored crosses) fall in the most strongly populated region of the plot,
with host masses in the range $10^{13.4-14.1}$ M$_\odot$ and pericenter distances below 120 kpc. Thus they can be
viewed as representative of a typical bar-like galaxy formed in a cluster via a tidal interaction with a
BCG.

As discussed in Section~2 using the examples of six selected galaxies, the strong mass loss during the
pericenter passages (tidal stripping) is accompanied by the loss of rotation (tidal stirring). In general, tidal
stirring transforms rotating disks into pressure-supported spheroids, but an intermediate stage of this process can
involve changing the orbits of the stars into more radial ones characteristic of bars, as is the case of the sample of
galaxies studied here. In Fig.~\ref{properties}c I plot the ratio of the final (present) rotation parameter $f$ to its
maximum value attained during a galaxy's history versus the ratio of the final dark matter content to the maximum dark
mass the galaxy ever possessed. As before, the colored crosses show the measurements for the selected six galaxies. As
expected, there is substantial correlation between the two parameters (with the correlation coefficient of 0.52),
although not very tight. In particular, five of the selected galaxies have $f (z=0)/f_{\rm max}$ between 0.1 and 0.2,
and only one (ID177727) retains more (0.38) of its maximum rotation support. At the same time, all the galaxies have
been heavily stripped and preserve less than 8\% of their maximum dark matter content.

The fraction of the dark matter mass remaining in the galaxy is a good proxy for the amount of tidal forcing the galaxy
experienced during its lifetime. It is therefore interesting to check if the strength of the bar at the end of the
evolution depends on the efficiency of the stripping, that is reflects to any extent the total tidal forcing the galaxy
underwent. To answer this question, in Fig.~\ref{properties}d I plot the present strength of the bar, as measured by
the maximum of the bar mode $A_{2, \rm{max}}$, as a function of the fraction of dark matter mass remaining in the
galaxy. The colored crosses again show the values for the selected six galaxies, but slightly different from those one
can read from Fig.~\ref{a2profiles}, because there I plotted the $A_2$ profiles at the last apocenter of the orbits
while here they are taken at $z=0$. The correlation between the two quantities is very weak (with the correlation
coefficient of only 0.05). In particular, the selected galaxies cover a wide range of $A_{2, \rm{max}}$, between 0.51
and 0.66, although they all fall in a rather narrow range of the remaining dark matter fractions.

This behavior can be easily understood by referring to the subsequent history of the galaxies. In the
previous sections it was shown that after bar formation the bar strength, as measured by the bar mode $A_2$ (within two
stellar half-mass radii, $2 r_{1/2}$), can increase, decrease or remain at the same level during the later evolution
(for example in the galaxies shown in Fig.~\ref{evolution1}). As discussed in detail by \citet{Lokas2014},
the fate of the bar at the subsequent pericenter passages is determined not only by the host mass and pericenter
distance, but also by the angle between the bar and the tidal force. The torque of the tidal force acting in the same
direction as the bar is rotating can speed up and weaken the bar, while the torque oriented opposite will have the
opposite effect and make the bar stronger. This effect is rather subtle and cannot be studied in detail here because of
insufficient number of simulation outputs that prohibit exact measurements of the angles just before the interaction.
However, the effect of the phenomenon can be discerned in the behavior of the bar mode in time, as shown in
Figs.~\ref{evolution1}, \ref{evolution2} and \ref{strongbars}.

\section{Discussion}

In this paper I studied bar-like galaxies from the IllustrisTNG simulations that were tidally induced in cluster
environments. These 77 objects comprise a subsample of 277 bar-like galaxies of various origin described in
\citet{Lokas2021}. I presented the six most convincing examples of such galaxies, in which the bar was induced after an
interaction with a progenitor of a massive BCG. I also described the properties of the bars in these galaxies in more
detail.

Then I discussed a few examples with more complicated history, involving multiple interactions and mergers, as
well as cases with particularly strong increase of the bar strength, not necessarily associated with the moment of bar
formation. In spite of these difficulties in the interpretation, the general picture emerging from this study is that
tidal interactions with massive galaxies in the cluster, most often progenitors of BCGs, can indeed induce bars in
galaxies, which confirms in the cosmological context the conclusions of earlier studies based on controlled simulations
\citep{Lokas2016, Kwak2017} and generalizes the results obtained previously for the single most massive cluster of
IllustrisTNG100 \citep{Lokas2020b}.

Finally, I studied the properties of the sample as a whole and found that the time of bar formation is very
strongly correlated with the time of one of the pericenter passages. The mass loss and rotation loss are also to some
extent correlated, while the final strength of the bar does not really depend on the amount of tidal force experienced
by the galaxy, as approximated by the mass loss. The reasons for this behavior include the fact that the progenitors of
the galaxies are different and therefore some are more and some are less susceptible to bar formation by themselves.
In addition, the evolution of the bar at subsequent pericenters is different depending on the orientation of the bar
with respect to the tidal force just before the interaction, which is a random combination of the bar pattern speed
and the velocity of the galaxy on the orbit. The tidal torque can increase or decrease the strength of the bar, and I
have presented numerous examples of both occurring in the analyzed galaxies.

The tidally induced bar-like galaxies studied here form with a variety of pericenter distances and for a wide range of
host masses, but a larger pericenter usually requires a more massive host, as expected from the nature of the tidal
force. Surprisingly, the configurations leading to the formation of the bar do not show any preference for prograde
orbits. The angles between the galaxy's internal angular momentum and the orbital one cover a wide range of
mostly intermediate values, with very few low (prograde) and high (retrograde) angles. This is unexpected since
controlled simulations clearly show that prograde orientations more often lead to the formation of tidally induced bars
\citep{Lokas2015, Lokas2018}. The only explanation I can think of is that the tidal field around the interacting galaxy
in the cluster environment is so complicated that the dependence on the angle of interaction even with the dominant
perturber is not the key factor. It seems that the host mass and the pericenter distance that contribute to the pure
tidal force are more important.

An important difference between the cosmological simulations analyzed here and the controlled simulations used to study
the phenomenon before is that the latter were very highly idealized and usually did not contain gas. The gas stripping
during the pericenter passages is inherent in all cases of tidal evolution in clusters. For the clearest cases
of interactions with BCGs (Figs.\ref{evolution1}-\ref{evolution2}), the gas is usually lost at the second pericenter
and already strongly depleted at the first, so the bar-forming galaxy is already significantly devoid of gas.
Indeed, as discussed in \citet{Lokas2021}, the tidally induced bars studied here (of class A in the previous paper),
had gas fractions within $2 r_{1/2}$ in the range 0-0.4 (60 galaxies had this fraction below 0.1, nine in the range
0.1-0.2, seven in the range 0.2-0.3 and only one at the level of 0.4) at the time of bar formation. Since we know that
lower gas fractions make galaxies more susceptible to bar formation \citep{Shlosman1993, Athanassoula2013, Lokas2020a},
this low gas content must have played a role in how effectively the bars were induced.

The times of bar formation for the sample studied here are very widely spread, from $t = 2.2$ to 12.1 Gyr. The
distribution peaks around 8-9 Gyr, when the formation of clusters and their massive BCGs is already under way, but
there are many cases of much earlier tidal bar formation. The extreme example studied here has the bar formed at 2.2
($z = 2.9$) via a very close passage near a massive galaxy, a progenitor of a BCG. This galaxy could be a candidate for
an early bar, one of the objects that are notoriously missing in cosmological simulations, but are being discovered in
observations \citep{Costantin2023, Amvrosiadis2025}. It is discussed in detail in a separate paper
\citep{Lokas2025}.

The bar-like galaxies studied here differ from the classical bars embedded in disks because in their structure the
bar dominates the whole stellar component. When viewed face-on, such objects would be most often classified in
observations as S0s, many of which have pronounced bars and rather faint disks, and are often found in galaxy clusters
\citep{Baillard2011}. A good example of those is NGC 4479, a member of the Virgo cluster. They would be much
more difficult to identify in observations when viewed not exactly face-on and could thus appear as ellipticals for
other lines of sight. Their prolate shape would be seen as elliptical in 2D projection and they should possess some
remnant rotation characteristic of bars. It may be difficult to confirm their prolate shape from such data since the
intrinsic 3D shapes of ellipticals have been persistently difficult to infer \citep{Weijmans2014, Bassett2019},
although recent studies appear to have made some progress in this area. Interestingly, it has been found that a larger
fraction of galaxies can be prolate rather than oblate than was previously believed, especially at high redshifts
\citep{Pandya2024}.

\begin{acknowledgements}
I am grateful to the anonymous reviewer for useful comments and to the IllustrisTNG team for making their
simulations publicly available. Computations for this work have been performed using the computer cluster at the
Nicolaus Copernicus Astronomical Center of the Polish Academy of Sciences (CAMK PAN).
\end{acknowledgements}

\end{document}